\documentclass[aps,twocolumn,amsmath,amssymb,superscriptaddress,prl
]{revtex4}

\usepackage{physics}
\usepackage{graphicx}% Include figure files
\usepackage{dcolumn}% Align table columns on decimal point
\usepackage{bm}% bold math
\usepackage{xcolor}

\usepackage{amsmath,amssymb}
\usepackage[normalem]{ulem}

\begin{document}
%\preprint{APS/123-QED}

\title{Theory of spin center sensing of diffusion and other surface electric dynamics}%\blue{and surface correlated phenomena}}% Force line breaks with \\

\author{Denis R. Candido}
 \email{denisricardocandido@gmail.com}
\affiliation{Department of Physics and Astronomy, University of Iowa, Iowa City, Iowa 52242,
USA}
%\affiliation{Pritzker School of Molecular Engineering, University of Chicago, Chicago, Illinois 60637, USA}

\author{Michael E. Flatt\'e}%
 \email{michael_flatte@mailaps.org}
\affiliation{Department of Physics and Astronomy, University of Iowa, Iowa City, Iowa 52242,
USA}
%\affiliation{Pritzker School of Molecular Engineering, University of Chicago, Chicago, Illinois 60637, USA}
\affiliation{Department of Applied Physics, Eindhoven University of Technology, P.O. Box 513, 5600 MB, Eindhoven, The Netherlands}

\date{\today}

%There must be an abstract of no more than 600 characters
\begin{abstract} 
Surface electric dynamics influences the quantum coherence of near-surface spin centers through spatial and temporal fluctuations of the surface charge density and the electrostatic potential at the crystal surface. 
%For a sub-surface spin center, t
The power law  of the electric noise's spectral density dependence on spin center depth is not solely determined by whether the surface charges fluctuate as monopoles or dipoles. The power instead depends sensitively on the structure of the surface charge and surface potential correlation functions.
For surface dynamics originating from diffusion, the spin center's relaxation and decoherence times exhibit a temporal crossover  near the correlation time of the fluctuators and thus  provide a quantitative fingerprint of the diffusive behavior of charged particles at surfaces. %characteristics?
\end{abstract}

\maketitle

{\it Introduction.---} Defects in solids possessing spin are promising solid-state qubits whose long spin coherence times permit optical  initialization and readout~\cite{Awschalom2002,anderson2021five}. The energy levels of spin-1  centers lacking inversion symmetry, such as NV centers in diamond, shift in response to electric fields to first order, motivating their application to quantum sensing and metrology of charge dynamics~\cite{taylor2008high,koenraad2011single,electric-magnetic1,dolde2014,schirhagl2014nitrogen,van2015nanometre,degen2017quantum,flebus2018quantum,casola2018probing,electricnoise3,PhysRevX.10.011003,lee2020nanoscale,rustagi2020,RevModPhys.92.015004}, including for nanoscale probes of many-body physics~\cite{magneticnoise7,PhysRevLett.121.023601,davis2021probing,dwyer2021probing}. Undesirable fluctuating electric and magnetic fields also couple to the spin\cite{electric-magnetic1,electric-magnetic2,electric-magnetic3,electric-magnetic4,electricnoise1,electricnoise2,electricnoise3,magneticnoise1,magneticnoise2,magneticnoise3,magneticnoise4,magneticnoise5,magneticnoise6,magneticnoise7}, and their effects increase near the surface due to trapped charges and imperfections.
The differing spatial dependence of fluctuating electric fields produced by {uncorrelated} monopole (point-like) and dipole charges at the surface, $E\propto |\textbf{r}|^{-2}$ and $E\propto |\textbf{r}|^{-3}$, %respectively, point-like (dipole) charges fluctuation 
 produce a surface noise spectral density $S_{E}(\omega)\propto d^{-2}$  and $S_E(\omega)\propto d^{-4}$~\cite{electricnoise1,electricnoise2,electricnoise3,magneticnoise6}, where $d$ is the spin center depth; this noise at a resonant frequency for spin transitions leads directly to spin relaxation and decoherence and reduces both the sensitivity to external fields and the optical contrast in readout. The rapid increase in noise near a surface conflicts with the desire to place the spin as close to the surface as possible to increase sensitivity to a signal originating from outside the material and to improve the sensor's spatial resolution. 
A major focus for quantum sensing, therefore, explores techniques to reduce these noise sources\cite{electricnoise1,electric-magnetic2,electric-magnetic3,anderson2019electrical,candido-pin}; similar goals help advance defect-based hybrid quantum systems~\cite{lukaprx,li2015hybrid,li2016hybrid,andrich2017long,lemonde2018phonon,flebus2019entangling,muhlherr2019magnetic,zou2020tuning,candido2020predicted,neumanprl2020,doi:10.1021/acs.jpcc.0c11536,solanki,fukami2021}, as the coupling of a spin to another excitation ({\it e.g.} magnon, phonon) also improves the closer the spin comes to the surface. These efforts depend on an accurate understanding of the noise and its depth-dependent scaling.

%Additionally, the current state of the art is that, due to the different scaling of 

%As a consequence, for shallow defects we have a strong increase of $S_E(\omega)$, which becomes a problem. \cyan{[Michael: this last sentence might now be a bit repetitive, see blue sentence below.]}
%For both cases, $S_{E}(\omega)\propto 1/\omega^{\alpha}$ with $\alpha=1-2$ is seen experimentally\cite{magneticnoise6}, in addition with a frequency independence of $S_{E}(\omega)$ for frequencies smaller than a critical frequency $\omega_c$~\cite{BarGill2012,magneticnoise4,magneticnoise5,magneticnoise6}.
%\begin{figure}[t!]
%\begin{center}
%\includegraphics[clip=true,width=.95\columnwidth]{fig0.pdf}
%\caption{Spin center as a quantum sensor of diffusion.} 
%\label{fig1}
%\end{center}
%\end{figure}

\begin{figure}[t!]
\begin{center}
\includegraphics[clip=true,width=1\columnwidth]{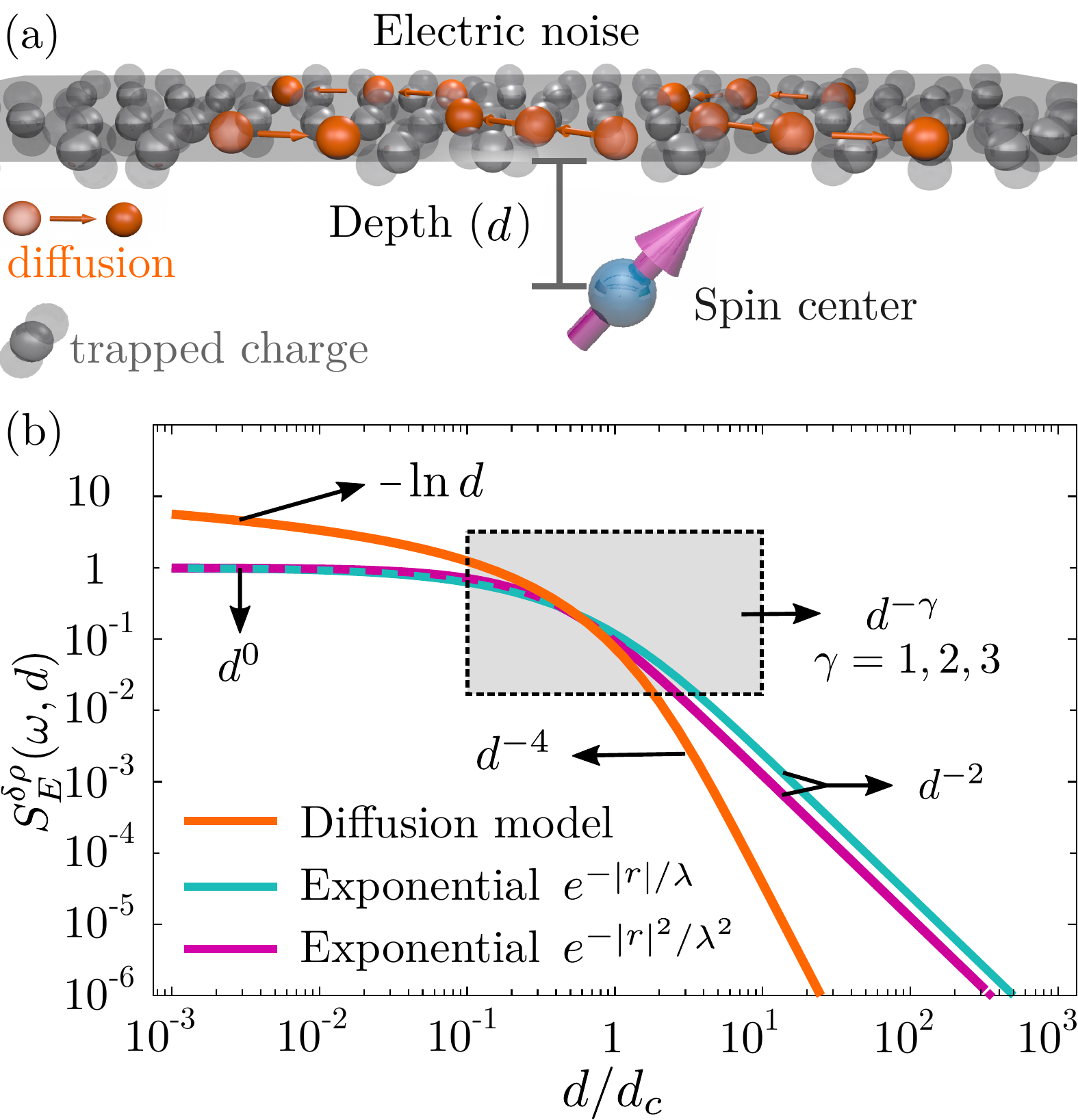}
\caption{(a) Electric surface noise (b) Electric noise spectral density $S^{\delta\rho}_E(\omega,d)$ for different correlation functions of the electronic surface density. For both exponential correlation functions %$e^{-|r|/\lambda}$ and $e^{-|r|^2/\lambda^2}$ 
%we obtain a constant depth dependence
$S^{\delta\rho}_E(\omega,d)\propto d^0$ for $d\ll d_c$ and $S^{\delta\rho}_E(\omega,d)\propto d^{-2}$  for $d\gg d_c$ ($d_c=\lambda$). For %a correlation function obeying the diffusion equation, 
diffusion  $S^{\delta\rho}_E(\omega,d)\propto -\ln d$  for $d\ll d_c$ and $d^{-4}$ for $d\gg d_c$ [$d_c=({\cal D}/\omega)^{1/2}$ where ${\cal D}$ is the diffusion constant]. Within the gray box  $S^{\delta\rho}_E(\omega,d)\propto d^{-\gamma}$ with $\gamma=1,2$ and $3$.} 
\label{fig1}
\end{center}
\end{figure}

Here we show that very different depth ($d$) dependences for the frequency ($\omega$) dependent noise spectral density $S_E(\omega)$ emerge when spatial {and time} correlations {are} considered; our general formalism  relies on the  two-point correlation functions of  the 
fluctuating point-like charges' surface density, $\left\langle \delta \rho(\textbf{r}',t')\delta \rho(\textbf{r},t)\right\rangle$, and the fluctuating electrostatic potential at the crystal surface, $\left\langle \delta \phi (\textbf{r}',t') \delta \phi (\textbf{r},t)\right\rangle$, {yielding}  $S_{E}^{\delta \rho}$ and $S_E^{\delta \phi}$, respectively.
For example, for diffusive motion of surface charges we find $S_{E}^{\delta \rho} \propto {d^{-4}}$ instead of the ${d^{-2}}$ {dependence} {found for uncorrelated} fluctuation of  point-like particle densities. We find $S_{E}^{\delta \rho}$ is {\it independent of depth}  when both (1) the fluctuating charge density spatial correlation falls off exponentially with distance  ({\it e.g.} from crystal  surface distortions and imperfections\cite{doi:10.1063/1.1698419,doi:10.1063/1.1722830}) and (2) the spin depth is much shorter than the correlation length. We further obtain $S_E^{\delta \phi}\propto {d^{-4}}$ for  exponentially decaying spatial correlations of the surface potential fluctuations. In general, the depth-dependent behavior of $S_E^{\delta \rho}$ and $S_E^{\delta \phi}$ strongly depends on the  detailed physics of the surface noise  phenomenon and {on} both $\left\langle \delta \rho(\textbf{r}',t')\delta \rho(\textbf{r},t)\right\rangle$ and  $\left\langle \delta \phi (\textbf{r}',t') \delta \phi (\textbf{r},t)\right\rangle$.
 $S_E^{\delta \rho}$ and $S_E^{\delta \phi}$ even exhibit  different depth dependences for the same mathematical form for both $\left\langle \delta \rho(\textbf{r}',t')\delta \rho(\textbf{r},t)\right\rangle$ and  $\left\langle \delta \phi (\textbf{r}',t') \delta \phi (\textbf{r},t)\right\rangle$.  These results suggest a new sensing methodology:  correlating surface-dependent physical phenomena with $d$ and $\omega$ dependent noise spectra detected by electric-field sensitive quantum sensors; for spin centers sensing diffusive behavior the diffusion constants and correlation times can be extracted from detailed time-dependent measurements of spin coherence. 
 %In general, our discovery brings another important factor to be studied and addressed in order to fully understand the defect coherence and dephasing time as a function of $d$.  
%Secondly, the theoretical understanding of our results establishes the possibility for using spin defects to sense diffusion %and drift behavior
%of particles, including the charges within our crystals, crystal surface, or even the charges of a liquid surrounding our crystal. As the defect's decoherence and dephasing rates are influenced by the electric field spectral noise density, we provide a way to extract the diffusion coefficient based on the response of the decoherence and dephasing time as a function of either depth or frequency.  Finally, we also show that the dephasing or decoherence as a function of time contains a crossover on their dependence, from which it is possible to extract the correlation time of diffusion phenomena. 

%Finally, we show that it is possible to use spin defects to sense indirectly the shot noise arising from the fluctuation of the charged particle density, $\left\langle n(\textbf{r}',t')n(\textbf{r},t)\right\rangle$.

%\ccyan{[We also develop a theory for the magnetic noise produced by the fluctuating charged particles, and similar results to the electric field noise are also found: Michael, I have not calculated this since I thought the paper is already strong enough.]}

{\it Fundamentals of electric noise at the diamond surface.---}
Hydrogen or oxygen termination shift the surface potential for diamond differently\cite{PhysRevB.73.085313,sussmann2009cvd,doi:10.1021/nl501927y,chou_gali_2017}; similar effects emerge from imperfections of crystal termination
%In addition, imperfections of the crystal termination are responsible for an extra creation of surface acceptors or donors~
\cite{https://doi.org/10.1002/admi.201801449} along with 
%and also a 
spatial fluctuations of the surface potential~\cite{PhysRev.104.619,PhysRevA.61.063418}. Interaction of these levels with trapped charges in the % For instance, these acceptors (donors) levels are occupied by the electrons (holes) arising from the 
nitrogen-doped diamonds that host NV centers result in the creation of an effective surface two-dimensional (2D)  hole or electron gas~\cite{KAWARADA1996205,PhysRevLett.85.3472,PhysRevB.68.041304,sussmann2009cvd,surfacediamondreview,https://doi.org/10.1002/admi.201801449}. These charges do not distribute uniformly and move around due to  thermal fluctuations of the charge's position, collision between charges, capture and ionization by donors and acceptors, and other processes. Moreover, as the read-out and initialization of the sensing spin requires laser illumination, the light increases these fluctuations. Thus the fluctuation of  charged particles and the local surface potential produce noise that is experienced by shallow spins, causing decoherence {and relaxation} of the  spin state. 

The influence of these temporal fluctuations of charge density and  surface potential on the spin center properties occurs via %$\delta\rho\left(\textbf{r},t\right)=\rho\left(\textbf{r},t\right)-\left\langle \rho\left(\textbf{r},t\right)\right\rangle=\rho\left(\textbf{r},t\right)- \rho_{eq}\left(\textbf{r}\right)$, 
 a temporally fluctuating electric field  {at the spin center position $\textbf{r}=\textbf{r}_{\rm{d}}$,} $\textbf{E}(\textbf{r}{_{\rm{d}}},t) = \left\langle  \textbf{E}(\textbf{r}{_{\rm{d}}}) \right\rangle + \delta \textbf{E}(\textbf{r} {_{\rm{d}}},t)$, which causes spin decoherence {and relaxation}~\cite{electric-magnetic1,electric-magnetic2,electric-magnetic3,electric-magnetic4,electricnoise1,electricnoise2,electricnoise3,magneticnoise1,magneticnoise2,magneticnoise3,magneticnoise4,magneticnoise5,magneticnoise6,magneticnoise7} and increases the photoluminescence linewidth~\cite{tamarat2006stark,anderson2019electrical,de2017stark,candido-pin}.
For Dirichlet boundary conditions 
\begin{widetext}
\begin{equation}
    {\textbf{E}\left(\textbf{r} {_{\rm{d}}},t\right)=\frac{1}{\varepsilon}\int_{\mathcal{V}} d\textbf{r}'\textbf{K}\left(\textbf{r} {_{\rm{d}}}-\textbf{r}'\right)\rho\left(\textbf{r}',t\right)+\oint_{\mathcal{S}}da'\phi\left(\textbf{r}'_{S},t\right)\frac{\partial\textbf{K}\left(\textbf{r} {_{\rm{d}}}-\textbf{r}'_{S}\right)}{\partial n'}},
\end{equation}
and
%\begin{align}
    %\left\langle \delta\textbf{E}\left(\textbf{r},t\right)\delta\textbf{E}\left({\textbf{r}},0\right)\right\rangle %&=\left(\frac{1}{\varepsilon}\right)^{2}\int_{\mathcal{V}}\int_{\mathcal{V}} %d\textbf{r}'d\textbf{r}''\textbf{K}\left(\textbf{r}-\textbf{r}'\right)\textbf{K}\left({\textbf{r}}-\textbf{r}''\right)\nonumber \times\left\langle %\delta\rho\left(\textbf{r}',t\right)\delta\rho\left(\textbf{r}'',0\right)\right\rangle \\
    %&+\int_{\mathcal{S}}\int_{\mathcal{S}}d{\mathcal{S}}'d{\mathcal{S}}''\frac{\partial\textbf{K}\left(\textbf{r}-\textbf{r}'\right)}{\partial %z'}\frac{\partial\textbf{K}\left({\textbf{r}}-\textbf{r}''\right)}{\partial z''}\left\langle %\delta\phi\left(\textbf{r}',t\right)\delta\phi\left(\textbf{r}'',0\right)\right\rangle. \label{E-corr-G}
%\end{align}
\begin{align}
    \left\langle \delta{E}_\nu \left(\textbf{r} {_{\rm{d}}},t\right)\delta{E}_\mu\left({\textbf{r} {_{\rm{d}}}},0\right)\right\rangle &=\left(\frac{1}{\varepsilon}\right)^{2}\int_{\mathcal{V}}\int_{\mathcal{V}} d\textbf{r}'d\textbf{r}''{K}_\mu\left(\textbf{r} {_{\rm{d}}}-\textbf{r}'\right){K}_\nu \left({\textbf{r}} {_{\rm{d}}}-\textbf{r}''\right)\nonumber \left\langle \delta\rho\left(\textbf{r}',t\right)\delta\rho\left(\textbf{r}'',0\right)\right\rangle \\
    &+\int_{\mathcal{S}}\int_{\mathcal{S}}d\textbf{r}'_\parallel d\textbf{r}''_\parallel\frac{\partial {K}_\mu (\textbf{r} {_{\rm{d}}}-\textbf{r}'_\parallel)}{\partial z}\frac{\partial {K}_\nu ({\textbf{r}} {_{\rm{d}}}-\textbf{r}''_\parallel)}{\partial z}
    \expval{\delta\phi(\textbf{r}'_\parallel,t)\delta\phi(\textbf{r}''_\parallel,0)} \label{E-corr-G}
\end{align}
\end{widetext}
for $\textbf{r}_S$ on ${\cal S}=\left\{ (x,y,z), \thinspace -L/2 \leq x,y \leq L/2,\thinspace \thinspace z=0 \right\}$,  $\textbf{r}_d$ in ${\cal V}= \left\{ (x,y,z), \thinspace -L/2 \leq x,y \leq L/2,\thinspace \thinspace z>0 \right\}$,  $n'$ the coordinate normal to ${\cal S}$  {and $\textbf{r}_{\rm{d}}=(0,0,-d)$}.
%where we have defined our surface as ${\mathcal{S}= \left\{ (x,y,z), \thinspace -L/2 \leq x,y \leq L/2,\thinspace \thinspace z=0 \right\}}$ and our volume as ${\mathcal{V}= \left\{ (x,y,z), \thinspace -L/2 \leq x,y \leq L/2,\thinspace \thinspace z>0 \right\}}$ and $\textbf{r}_S=\textbf{r}_\parallel=(x,y,0)$. 
{Here, the kernel $\textbf{K}(\textbf{r})$ and its derivative with respect to the surface normal ($z$) are expressed as a 2D Fourier transform (${\cal F}_{2D}^{\textbf{k}_{\parallel}} [\textbf{K}(\textbf{r})]= \int{d\textbf{r}_{\parallel}} \textbf{K}(\textbf{r}) e^{-i\textbf{k}_{\parallel}\cdot\textbf{r}_{\parallel}}$)
with respect to the surface coordinates (yielding the Weyl representation of the Green's function)}
\begin{align}
{\cal F}_{2D}^{\textbf{k}_{\parallel}}[K_{\nu}(\textbf{r})]
&=  \frac{1}{2}\left[i\left(\delta_{\nu,x}+\delta_{\nu,y}\right)k_{\nu}/\left|\textbf{k}_{\parallel}\right|+\delta_{\nu,z}\right]e^{-\left|\textbf{k}_{\parallel}\right|\left|z\right|},\nonumber \\
{\cal F}_{2D}^{\textbf{k}_{\parallel}}\left[\partial_{z}K_{\mu}\left(\textbf{r}\right)\right]&=\partial_z {\cal F}_{2D}^{\textbf{k}_{\parallel}}[K_{\mu}(\textbf{r})].\label{Weylrep}
\end{align}
%$\blue{{\textbf{K}(\textbf{r}-\textbf{r}'_{\parallel})=\left(2\pi\right)^{-2}\int{d\textbf{k}_{\parallel}}{\cal F}_{2D}^{\textbf{k}_{\parallel}}[\textbf{K}(\textbf{r}-\textbf{r}'_{\parallel})]e^{i\textbf{k}_{\parallel}\cdot\left(\textbf{r}{}_{\parallel}-\textbf{r}'_{\parallel}\right)}}}$
%The correlation of the {fluctuating} electric field [Eq.~(\ref{E-corr-G})] is determined by the correlation of  the fluctuating surface electrostatic potential (as derived in Refs.~\onlinecite{PhysRevA.80.031402} and~\onlinecite{PhysRevA.84.053425}) {\it but also} the fluctuating surface charge density, where $\left\langle ... \right\rangle $ is the average over the ensemble of fluctuations.
%charges and potentials.

%, $\left\langle \delta\rho\left(\textbf{r}',t\right)\delta\rho\left(\textbf{r}'',0\right)\right\rangle$, 
%$\left\langle \delta\phi(\textbf{r}'_\parallel,t)\delta\phi(\textbf{r}''_\parallel,0)\right\rangle$~

We restrict fluctuations of $\rho$ and $\phi$ 
%density and surface potential 
to occur only on $\cal{S}$, thus $\rho\left(\textbf{r},t\right)=q\delta(z) n\left(\textbf{r}_\parallel,t\right)$, with $q$ the fluctuator's charge.  We {also} assume {translational} symmetry of the correlation function for the {fluctuating surface} density and surface potential along the surface, ${\Pi_{\delta \rho}(\textbf{r}'_\parallel-\textbf{r}_\parallel,t)=q^2 \left\langle \delta n(\textbf{r}'_\parallel,t)\delta n\left(\textbf{r}_\parallel,0\right)\right\rangle}$ and ${\Pi_{\delta \phi}(\textbf{r}'_\parallel-\textbf{r}_\parallel,t)=\left\langle \delta\phi(\textbf{r}'_{\parallel},t)\delta\phi(\textbf{r}_\parallel,0)\right\rangle }$, respectively. 
%\begin{equation}
%    \left\langle \delta n(\textbf{r}'_\parallel,t)\delta n(\textbf{r}_\parallel,0)\right\rangle=\left\langle \delta n(\textbf{r}'_\parallel-\textbf{r}_\parallel,t)\delta n\left(0,0\right)\right\rangle
%\end{equation}
%\begin{equation}
%{\left\langle \delta\phi(\textbf{r}'_{\parallel},t)\delta\phi(\textbf{r}''_{\parallel},0)\right\rangle}=\left\langle \delta\phi(\textbf{r}'_{\parallel}-\textbf{r}''_{\parallel},t)\delta\phi(0,0)\right\rangle 
    %\left\langle \delta\phi(\textbf{r}'_{\parallel},t)\delta\phi(\textbf{r}''_{\parallel},0)\right\rangle
%\end{equation}
Finally, 
${S_{E_\nu}(\omega)=\int_{-\infty}^{\infty}dt\left\langle \delta E_{\nu}\left(\textbf{r},t\right)\delta E_{\nu}\left(\textbf{r},0\right)\right\rangle e^{-i\omega t}}$, yielding
%in order to obtain $S_{E_\mu}(\omega)$, we still need to perform the Fourier transformation from $t$ to $\omega$ space, i.e., $\left\langle \delta E_{\nu}\left(\textbf{r},t\right)\delta E_{{\nu}}\left(\textbf{r},0\right)\right\rangle =\int_{-\infty}^{\infty}{d\omega}/{(2\pi)}S_{E_{\nu}}(\omega)e^{i\omega t}$ and ${\cal A}\left(\textbf{r}_\parallel,t\right)=1/(2\pi)^3\int{d\textbf{k}_\parallel}\int{d\omega}e^{i(\omega t+\textbf{k}_\parallel \cdot\textbf{r}_\parallel)}{\cal F}_{2D}^{\textbf{k}_\parallel, \omega}[{\cal A}(\textbf{r}_\parallel,t)]$, yielding 
\begin{widetext}
\begin{align}
   S_{E_{\nu}}^{\delta \rho}\left(\omega\right)&=\left(\frac{1}{\varepsilon}\right)^{2}\int\frac{d\textbf{k}_{\parallel}}{\left(2\pi\right)^{2}}{\cal F}_{2D}^{\textbf{k}_{\parallel}}\left[K_{\nu}\left(\textbf{r}{_{\rm{d}}}\right)\right]{\cal F}_{2D}^{-\textbf{k}_{\parallel}}\left[K_{\nu}\left(\textbf{r}{_{\rm{d}}}\right)\right]{\cal F}_{2D}^{\textbf{k}_{\parallel},\omega}[\Pi_{\delta \rho}(\textbf{r}'_\parallel-\textbf{r}_\parallel,t)], \label{E-corr-G-omega0} \\ 
      S_{E_{\nu}}^{\delta \phi}\left(\omega\right)&=\int\frac{d\textbf{k}_{\parallel}}{\left(2\pi\right)^{2}}{\cal F}_{2D}^{\textbf{k}_{\parallel}}[\partial_{z}K_{\nu}(\textbf{r}{_{\rm{d}}})]{\cal F}_{2D}^{-{\textbf{k}}_{\parallel}}[\partial_{z}K_{\nu}(\textbf{r}{_{\rm{d}}})]{\cal F}_{2D}^{\textbf{k}_{\parallel},\omega}[\Pi_{\delta \phi}(\textbf{r}'_\parallel-\textbf{r}_\parallel,t) ] .\label{E-corr-G-omega}
\end{align} 
\end{widetext}
Equations~(\ref{Weylrep})-(\ref{E-corr-G-omega}) establish the relation between the electric spectral noise and the temporal and spatial Fourier transform of {both $\Pi_{\delta\rho}(\textbf{r}'_\parallel-\textbf{r}_\parallel,t)$ and $\Pi_{\delta\phi}(\textbf{r}'_\parallel-\textbf{r}_\parallel,t)$.} %the correlation of the {fluctuating} surface density and potential at different times and positions. \red{in terms of $\Pi_{\delta\rho}$ and $\Pi_{\delta\phi}$}

{\it Fluctuating surface density from diffusion ---} We assume {$ n\left(\textbf{r}_\parallel,t\right)$} satisfies 
\begin{equation}
\left(\frac{\partial}{\partial t}+\frac{1}{\tau}+\mu\textbf{E}_{\parallel}\cdot\nabla_\parallel-{\cal D}\nabla^{2}_\parallel\right){ n\left(\textbf{r}_\parallel,t\right) }=0, \label{diff-n}
\end{equation}
which includes %many physical processes of transport of particles including
diffusion, drift and a finite carrier lifetime. Accordingly, the {corresponding} Green's function satisfies 
\begin{equation}{
(\frac{\partial}{\partial t}+\frac{1}{\tau}+\mu\textbf{E}_{\parallel}\cdot\nabla_\parallel-{\cal D}\nabla^{2}_\parallel){\cal G}\left(\textbf{r}_\parallel,t\right)=\delta(\textbf{r}_\parallel)\delta(t)} \label{diff-n2}
\end{equation}
and
${{ \delta n(\textbf{r}_\parallel,t)} =\int d\textbf{r}'{\cal G}(\textbf{r}_\parallel-\textbf{r}'_\parallel,t) { \delta n(\textbf{r}'_\parallel,0)} }$, so  {${\Pi(\textbf{r}'_{\parallel}-\textbf{r}_{\parallel},t)=\int d\textbf{r}''_{\parallel}{\cal G}(\textbf{r}''_{\parallel}-\textbf{r}'_{\parallel},t)\Pi(\textbf{r}'_{\parallel}-\textbf{r}_{\parallel},0)} $.}
%\begin{widetext}
%\begin{equation}
%\Pi(\textbf{r}'_{\parallel}-\textbf{r}_{\parallel},t) = \left\langle \delta n(\textbf{r}'_{\parallel}-\textbf{r}_{\parallel},t)\delta n(0,0)\right\rangle  =\int d\textbf{r}''_{\parallel}{\cal G}(\textbf{r}''_{\parallel}-\textbf{r}'_{\parallel},t)
% \left\langle \delta n(\textbf{r}''_{\parallel}-\textbf{r}_{\parallel},0)\delta n(0,0)\right\rangle .
% \end{equation}
%\end{widetext}
%\begin{align}
%\left\langle \delta n(\textbf{r}'_{\parallel}-\textbf{r}_{\parallel},t)\delta n(0,0)\right\rangle  & =\int d\textbf{r}''_{\parallel}{\cal G}(\textbf{r}''_{\parallel}-\textbf{r}'_{\parallel},t)\nonumber \\
% & \times\left\langle \delta n(\textbf{r}''_{\parallel}-\textbf{r}_{\parallel},0)\delta n(0,0)\right\rangle  \label{corr-time-space}
%\end{align}
With this,  Eq.~(\ref{E-corr-G-omega0}) connects the noise's spectral density to both the electrostatic Green's function {Kernel} $\textbf{K}(\textbf{r}'-\textbf{r})$ %\left[K_{\nu}\left(\textbf{r}\right)\right]$
and the Green's function describing the dynamics of our charged particles, ${\cal G}$. %(\textbf{r}''_{\parallel}-\textbf{r}'_{\parallel},t)$. 
{For purely diffusive time evolution of the fluctuation correlator,}  ${\left\langle \delta n(\textbf{r}'_{\parallel},0)\delta n(\textbf{r}_{\parallel},0)\right\rangle = \delta(\textbf{r}'_{\parallel}-\textbf{r}_{\parallel})\left\langle  n(\textbf{r}_\parallel,0)\right\rangle}$\cite{van1992stochastic,nano11020358}, we find {$\Pi(\textbf{r}'_{\parallel}-\textbf{r}_{\parallel},t) = {\cal G}(\textbf{r}'_{\parallel}-\textbf{r}_{\parallel},t)\left\langle n\left(\textbf{r}_\parallel,0\right)\right\rangle$ and ${\cal F}_{2D}^{\textbf{k}_{\parallel},\omega}[ \Pi(\textbf{r}'_{\parallel}-\textbf{r}_{\parallel},t) ]={\cal F}_{2D}^{\textbf{k}_{\parallel},\omega}\left[{\cal G}\left(\textbf{r}{}_{\parallel},t\right)\right] n_S$ with $n_S=\left\langle n\left(\textbf{r}{}_{\parallel},0\right)\right\rangle$ for uniform density, and ${\cal F}_{2D}^{\textbf{k}_\parallel,\omega}[{\cal G}(\textbf{r}_\parallel,t)]=({i\omega+1/\tau+i\mu\textbf{E}_\parallel \cdot\textbf{k}_\parallel+{\cal D}\textbf{k}_\parallel^{2}})^{-1} $ following} from Eq.~(\ref{diff-n2}). 
%\begin{equation}
%    \xout{\left\langle \delta n(\textbf{r}'_{\parallel},t)\delta n\left(\textbf{r}_\parallel,0\right)\right\rangle ={\cal G}(\textbf{r}'_{\parallel}-\textbf{r}_{\parallel},t)\left\langle n\left(\textbf{r}_\parallel,0\right)\right\rangle,}
%\end{equation}
%\begin{equation}
% \xout{{\cal F}_{2D}^{\textbf{k}_{\parallel},\omega}\left[\left\langle \delta n\left(\textbf{r}{}_{\parallel},t\right)\delta n\left(0,0\right)\right\rangle \right]={\cal F}_{2D}^{\textbf{k}_{\parallel},\omega}\left[{\cal G}\left(\textbf{r}{}_{\parallel},t\right)\right]\left\langle n\left(0,0\right)\right\rangle}
% \end{equation}
%\begin{equation}
 %   {\cal F}_{2D}^{\textbf{k}_{\parallel},\omega}\left[\left\langle \delta n\left(\textbf{r}{}_{\parallel},t\right)\delta n\left(0,0\right)\right\rangle \right]&={\cal F}_{2D}^{\textbf{k}_{\parallel},\omega}\left[{\cal G}\left(\textbf{r}{}_{\parallel},t\right)\right]\left\langle n\left(0,0\right)\right\rangle .
%\end{equation}
%and 
%\begin{equation}
%\xout{{\cal F}_{2D}^{\textbf{k}_\parallel,\omega}[{\cal G}(\textbf{r}_\parallel,t)]=({i\omega+1/\tau+i\mu\textbf{E}_\parallel %\cdot\textbf{k}_\parallel+{\cal D}\textbf{k}_\parallel^{2}})^{-1}  ,  }
%\end{equation}
 
\begin{figure*}
  \includegraphics[width=\textwidth,height=10cm]{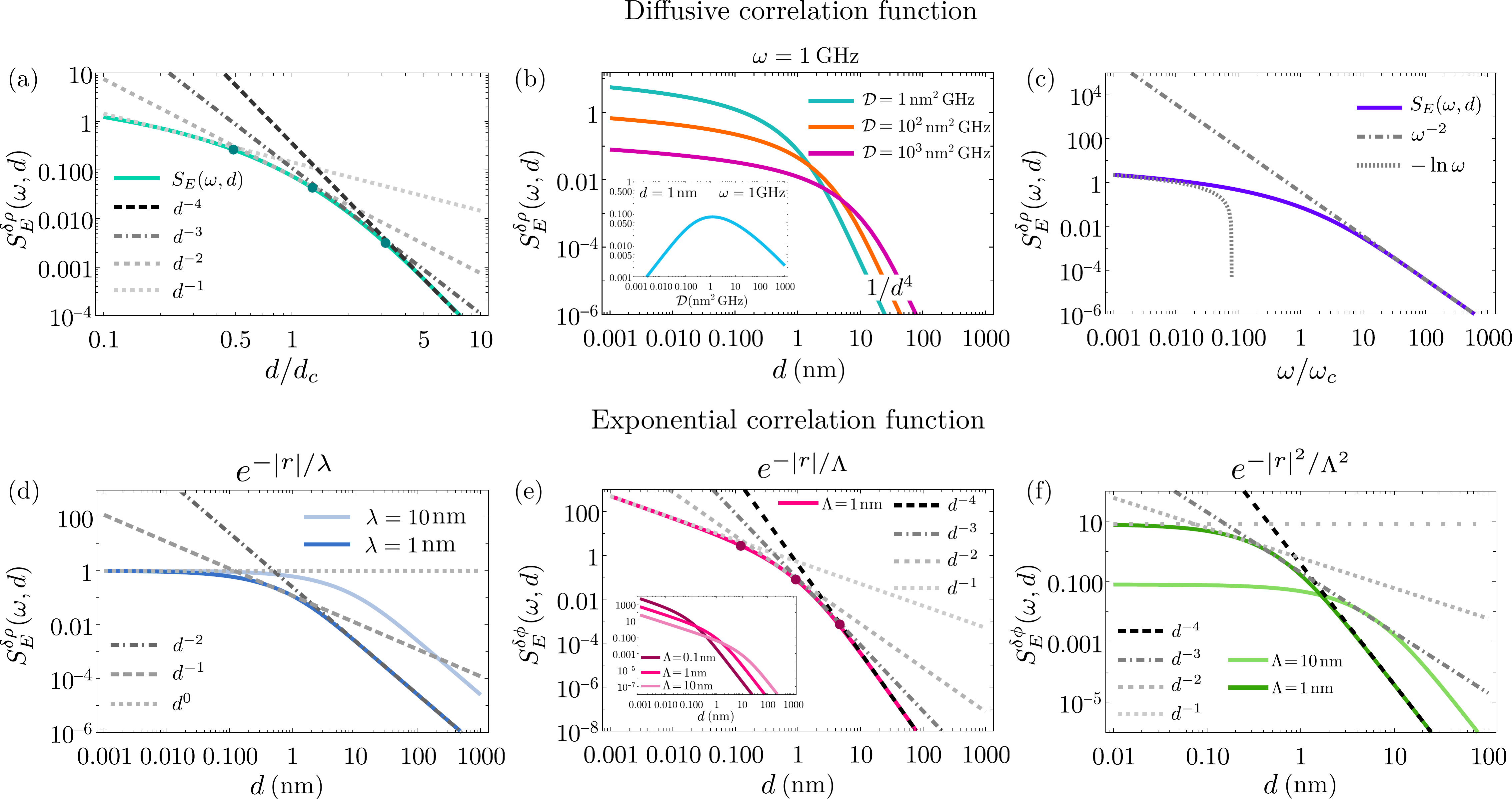}
  \caption{(a) $S_{E}^{\delta \rho}(\omega)$ [Eq.~(\ref{SEx-rho})] as a function of $d/d_c$ ($d_c=[{{\cal D}/\omega}]^{1/2}$). The dashed and dotted lines are guides for the eyes for ${d^{-\gamma}}$  with $\gamma=1,2,3$ and $4$, and the green circles show the depths where the $\gamma$ changes. (b) Same as (a) for different ${\cal D}=1,10,100$ (nm$^2$ GHz) and {$d_{c}=1$~nm}. Inset shows the $S_{E}^{\delta \rho}(\omega)$ [Eq.~(\ref{SEx-rho})] dependence on ${\cal D}$ for $d=1$~nm and $\omega=1$~GHz. (c) $S_{E}^{\delta \rho}(\omega)$ [Eq.~(\ref{SEx-rho})] as a function of $\omega/\omega_c$ ($\omega_c={\cal D}/d^2$), where dotted and dashed lines correspond to the asymptotic analytical forms within Eq.~(\ref{SE-diff}). (d) $S_{E}^{\delta \rho}(\omega)$ [Eq.~(\ref{exp-one})] as a function of $d$ for $\lambda={1}$ and {$10$}~nm, with dotted, dot-dashed and dashed lines corresponding to the asymptotic analytical forms [Eq.~(\ref{exp-one-asym})]. (e) Same as (d) for $S_{E}^{\delta \phi}(\omega)$, Eq.~(\ref{corr-exp-phi1}), with inset showing the $d$ dependence for different $\Lambda$ values. (f) Same as (e) for $S_{E}^{\delta \phi}(\omega)$, Eq.~(\ref{corr-exp-phi2}).  }
  \label{fig2}
\end{figure*}
We now use this equation within Eq.~{(\ref{E-corr-G-omega0})} assuming first $\textbf{E}_{\parallel}={\textbf{0}}$ and $1/\tau\rightarrow0$, i.e., purely diffusive motion, which for $\mu=x,y$ yields
\begin{align}
S_{E_{\mu}}^{\delta \rho}\left(\omega\right)=\left(\frac{q}{4\pi\varepsilon}\right)^{2}{n_{{\cal S}}\pi}\times\frac{2{\cal D}}{\omega^{2}d^{4}}\int_{0}^{\infty}d\epsilon\thinspace\frac{\epsilon^{3}e^{-2\epsilon}}{1+\frac{{\cal D}^{2}}{\omega^{2}d^{4}}\epsilon^{4}}, \label{SEx-rho} 
\end{align}
where $\epsilon$ is a dimensionless variable and ${S_{E_{z}}{/2}=S_{E_{x}}\equiv S_{E}}$. We cannot decouple the temporal part of the spectral noise from the spatial part. However, there are two limits, dictated by the value of $d^2 \omega/{\cal D}$, in which  Eq.~({\ref{SEx-rho}}) can be solved analytically, yielding
\begin{align}
S_{{E}}^{\delta\rho}\left(\omega\right)\approx\begin{cases}
{B_0} ({3{\cal D}}/{4d^{4}\omega^{2}}) & \thinspace d^2 \omega/{\cal D}\gg 1\\ %\label{SEx-app}
B_0 \left[\ln\left({{\cal D}}/{4\omega d^{2}}\right)+2\Gamma\right]/{{\cal D}} & d^2 \omega/{\cal D}\ll 1 
\end{cases}\label{SE-diff}
\end{align}
with $B_0=\left({q}/{4\pi\varepsilon}\right)^{2}n_{S}\pi$ and the Euler-Gamma number $\Gamma=0.577$. In Fig.~\ref{fig1}(a) we plot $S_E^{\delta \rho}$ [Eq.~(\ref{SEx-rho})] as a function of $d/d_c$ with $d_{c}=({{\cal D}/\omega})^{1/2}$. Although $S_{E}^{\delta \rho} \propto d^{-4}$ for $d\gg d_c$, we obtain a $-\ln{d}$ dependence for $d \ll d_c$, with the intermediate {gray} region { $0.1  \lesssim d/d_c \lesssim 10$}  yielding $d^{-\gamma}$ (with $\gamma=1,2$ and $3$) shown in Fig.~\ref{fig2}(a).

A $d^{-4}$ depth dependence of the spin coherence time reported in the literature was attributed to the dipole character of the fluctuators~\cite{magneticnoise4,magneticnoise5,magneticnoise6,candido-spin1surface}. Our results show that instead, this depth dependence can also be obtained from point-like charges that diffuse.
% Diffusive motion of charges occurs on crystal surfaces in which crystal imperfections create traps. 
 In Fig.~\ref{fig2}(b) we plot $S_{E}^{\delta \rho}$ [Eq.~(\ref{SEx-rho})] as a function of $d$ for different diffusion coefficients. We note that for  {$d\ll 1$ nm} faster diffusion (larger $\cal{D}$) suppresses noise, similar to  motional narrowing~\cite{WAnderson1954,Berthelot2006}. However{,} for $d \gg 1$~nm we obtain the opposite behavior, and a plot of $S_E^{\delta \rho}$ as a function of ${\cal D}$ shows a maximum at ${\cal D}\approx 1$ nm$^2$ GHz  for $\omega=1$~GHz and $d=1$~nm [Inset Fig.~\ref{fig2}(b)]. We plot in Fig.~\ref{fig2}(c) $S_E^{\delta \rho}$ as a function of $\omega$ where we see the crossover between $\omega^{-2}$ and $-\ln{\omega}$ at $\omega_c = {\cal D}/d^2$.

{\it Fluctuating surface density with exponential spatial correlation---}
An exponential correlation function of  the fluctuating surface  density, i.e.,
${\Pi(\textbf{r}'_{\parallel}-\textbf{r}_{\parallel},t) }	=\chi\left(t\right)( n_S/A )e^{-{\left|\textbf{r}'_{\parallel}-\textbf{r}_{\parallel}\right|^{\ell}}/\lambda^{\ell}}$ with ${\ell}>0\in {\mathbb Z}$ and $A$ the surface area, is  common in crystals with inhomogeneous  surfaces, and it can be obtained theoretically assuming random crystal surface distortions and imperfections~\cite{doi:10.1063/1.1698419,doi:10.1063/1.1722830}. This behavior has also been experimentally observed with X-ray scattering from crystal surfaces~\cite{doi:10.1063/1.1698419,doi:10.1063/1.1722830,https://doi.org/10.1002/adma.201401094,PhysRevApplied.16.054032}, which directly measures the degree of spatial correlation between  fluctuations of the electronic density. Accordingly, {${{\cal F}_{2D}^{\textbf{k}_{\parallel}}[\Pi(\textbf{r}'_{\parallel}-\textbf{r}_{\parallel},t)]=\chi\left(t\right)( n_S/A ){2\pi\lambda^{2}}/{(1+\lambda^{2}\textbf{k}_{\parallel}^{2})^{3/2}}}$ and ${{\cal F}_{2D}^{\textbf{k}_{\parallel}}[\Pi(\textbf{r}'_{\parallel}-\textbf{r}_{\parallel},t)]=\chi\left(t\right)( n_S/A ) \pi\lambda^{2}\exp (-{\lambda^{2}\textbf{k}_{\parallel}^2}/{4})}$}
%\begin{widetext}
%\begin{equation}
%\xout{{{\cal F}_{2D}^{\textbf{k}_{\parallel}}\left[\left\langle \delta n\left(\textbf{r}{}_{\parallel},t\right)\delta n\left(0,0\right)\right\rangle %\right]=\chi\left(t\right)( n_S/A ){2\pi\lambda^{2}}/{(1+\lambda^{2}\textbf{k}_{\parallel}^{2})^{3/2}}}}
%\end{equation}
%\begin{equation}
%\xout{{{\cal F}_{2D}^{\textbf{k}_{\parallel}}\left[\left\langle \delta n\left(\textbf{r}{}_{\parallel},t\right)\delta n\left(0,0\right)\right\rangle %\right]=\chi\left(t\right)( n_S/A ) \pi\lambda^{2}e^{-{\lambda^{2}\textbf{k}_{\parallel}^2}/{4}}}}
%\end{equation} 
for ${\ell}=1$ and ${\ell}=2$, respectively. Then from Eq.~{(\ref{E-corr-G-omega})}
\begin{widetext} 
\begin{align}
   { S_{{E}}^{{\delta \rho}, {\ell}=1}\left(\omega\right)}{}&=\left(\frac{q}{\varepsilon}\right)^{2}\frac{n_{S}}{8A}\frac{{\cal F}_{2D}^{\omega}\left[\chi\left(t\right)\right]}{\left(d/\lambda\right)^{2}}\int_{0}^{\infty}d\epsilon\frac{\epsilon e^{-2\epsilon}}{[1+(\lambda/d)^{2}\epsilon^{2}]^{3/2}},\label{exp-one}\\
    {S_{{E}}^{{\delta \rho},{\ell}=2}\left(\omega\right)}{}&=\left(\frac{q}{\varepsilon}\right)^{2}\frac{n_{S}}{8A}\frac{{\cal F}_{2D}^{\omega}\left[\chi\left(t\right)\right]}{\left(d/\lambda\right)^{2}}\int_{0}^{\infty}d\epsilon\thinspace\epsilon \exp{{-2\epsilon-\frac{\epsilon^{2}}{4(d/\lambda)^2}}} .\label{exp-two}
\end{align}
\end{widetext}
Both exponential correlation functions depend on the critical length  $d_c=\lambda$, and asymptotically
\begin{align}
    S_{{E}}^{{\delta \rho},{\ell}=1}\left(\omega\right)=\begin{cases}
B_{1}{\cal F}_{2D}^{\omega}\left[\chi\left(t\right)\right] & d\ll d_c, \\
B_{1}{\cal F}_{2D}^{\omega}\left[\chi\left(t\right)\right]/[{4\left(d/d_c\right)^{2}}] & \thinspace d\gg d_c,\\
\end{cases}  \label{exp-one-asym}
\end{align}
\begin{align}
    S_{{E}}^{{\delta \rho},{\ell}=2}\left(\omega\right)=\begin{cases}
B_{1}{\cal F}_{2D}^{\omega}\left[\chi\left(t\right)\right] & d\ll d_{c},\\
B_{1}{\cal F}_{2D}^{\omega}\left[\chi\left(t\right)\right]/[{8\left(d/d_{c}\right)^{2}}] & \thinspace d\gg d_{c},
\end{cases} \label{exp-two-asym}
\end{align}
with $B_{1}=\left({q}/{\varepsilon}\right)^{2}n_{S}/(8A)$. We see that both exponential correlation functions produce similar asymptotic behaviors, with a  $d^{-2}$ depth dependence for $d\gg d_c=\lambda$. {This depth dependence can also be obtained by taking the limit ${\lambda\rightarrow0}$ leading to ${\left\langle \delta n(\textbf{r}'_{\parallel},t)\delta n(\textbf{r}_{\parallel},0)\right\rangle \propto \delta(\textbf{r}'_{\parallel}-\textbf{r}_{\parallel})f(t)}$. This represents a situation where uncorrelated fluctuations add incoherently. % through their intensities. 
%As a consequence, this shows that 
Thus a reported $S_E\propto d^{-2}$~\cite{electricnoise1,electricnoise2,electricnoise3,magneticnoise6}  {can only be accurately interpreted as due to the}  fluctuation of point-like charges {for}  correlation lengths much smaller than the defect depth. As $d\approx 10$~nm, this implies that $\lambda\lesssim 1$~nm.} Compared to Eq.~(\ref{SE-diff}), Eqs.~(\ref{exp-one-asym}) and (\ref{exp-two-asym}) show that different classes of {surface} {density} spatial correlation function yield different depth dependences. In Fig.~\ref{fig2}(d) we plot Eq.~(\ref{exp-one}) as a function of $d$ for $\lambda=1$ and ${10}$ nm, together with its asymptotic functions given by Eq.~(\ref{exp-one-asym}). Surprisingly, there is a plateau of $S_E^{\delta \rho}$ for depth much shorter than $d_c=\lambda$, with nominal value independent of the power of the exponential correlation [See Fig.~\ref{fig1}(b)]. {This independence of depth  can be understood through ${\lambda\rightarrow \infty}$, which produces ${\left\langle \delta n(\textbf{r}'_{\parallel},t)\delta n(\textbf{r}_{\parallel},0)\right\rangle \propto f(t)}$, i.e., the fluctuations are strongly correlated and add coherently.} We also see that for intermediate depth values, $d\approx d_c$, $S_E^{\delta \phi}$ scales with ${d^{-1}}$.

{\it Effect of the fluctuating surface potential.---} 
The fluctuation of the electrostatic
potential cannot be described by the diffusive model Eq.~(\ref{diff-n}).
Alternatively, we assume that the surface is composed of  plaquettes~\cite{PhysRevA.61.063418,PhysRevA.80.031402} with varying local  potential, $\phi_{i}\left(t\right)$ and $\left\langle \phi_{i}(t)\right\rangle =\phi_{eq}$. Moreover, we
assume $\delta\phi\left(\textbf{r}_{\parallel},t\right)=\sum_{{i=1}}^{N}\delta\phi_{i}\left(t\right)\Theta_i(\textbf{r}_{\parallel})$
where $\Theta_i(\textbf{r}_{\parallel})$ is the square function associated with the plaquette $i$, i.e., it assumes value $1$ $(0)$ inside (outside) plaquette $i$, centered at $\textbf{r}_{\parallel}$~\cite{PhysRevA.61.063418,PhysRevA.80.031402}, and with $\int_{\cal S} d\textbf{r}_{\parallel}\sum{}_{i}\Theta_i(\textbf{r}_{\parallel})=A$. Using ${\left\langle \delta\phi_{i}\left(t\right)\delta\phi_{j}\left(0\right)\right\rangle =\delta_{ij}f_{\phi}\left(t\right)}$,
$\sum{}_{i}\left\langle \Theta_i(\textbf{r}_{\parallel})\Theta_j(\textbf{r}'_{\parallel})\right\rangle=N\left\langle \Theta_i(\textbf{r}_{\parallel})\Theta_i(\textbf{r}'_{\parallel})\right\rangle $ due to the randomness of $\Theta$, where $N$ is the {total} number of plaquettes. Assuming  spatial transitional symmetry of the correlation function of the plaquette positions, $\left\langle \Theta_i(\textbf{r}_{\parallel})\Theta_i(\textbf{r}'_{\parallel})\right\rangle = e^{-|\textbf{r}_{\parallel}-\textbf{r}'_{\parallel}|^{\ell}/\Lambda^{\ell}}$~\cite{PhysRevA.80.031402, PhysRevA.84.053425}. For ${\ell}=1$ and ${\ell}=2$, we obtain 
\begin{align}
    S_{E_{x}}^{\delta \phi ,{\ell}=1}\left(\omega\right)&={\cal F}_{2D}^{\omega}\left[f_{\phi}\left(t\right)\right]N\frac{\Lambda^{2}}{8d^{4}}\int_{0}^{\infty}d\epsilon\frac{\epsilon^{3}e^{-2\epsilon}}{[1+(\Lambda/d)^{2}\epsilon^{2}]^{3/2}}, \label{corr-exp-phi1}\\
   S_{E_{x}}^{{\delta \phi}, {\ell}=2}\left(\omega\right)&={\cal F}_{2D}^{\omega}\left[f_{\phi}\left(t\right)\right]N\frac{\Lambda^{2}}{4d^{4}}\int_{0}^{\infty}d\epsilon\epsilon^{3}e^{-2\epsilon}e^{-\left(\Lambda/d\right)^{2}\epsilon^{2}/4}, \label{corr-exp-phi2}
\end{align}
where Eq.~(\ref{corr-exp-phi1}) appears in Ref.~\onlinecite{PhysRevA.80.031402}. The above integrals exhibit asymptotic analytical behavior given by 
\begin{align}
S_{E_{x}}^{\delta \phi ,{\ell}=1}\left(\omega\right)&=\begin{cases}
{\cal F}_{2D}^{\omega}\left[f_{\phi}\left(t\right)\right]({N}/{16 d\Lambda}) & d\ll\Lambda,\\
{\cal F}_{2D}^{\omega}\left[f_{\phi}\left(t\right)\right]({3N\Lambda^{2}}/{64d^{4}}) & d\gg\Lambda,
\end{cases}     \label{corr-exp-phi1-asymp}\\
\vspace{5mm}
S_{E_{x}}^{\delta \phi,{\ell}=2}\left(\omega\right)&=\begin{cases}
{\cal F}_{2D}^{\omega}\left[f_{\phi}\left(t\right)\right]({N}/{\Lambda^{2}}) & d\ll\Lambda,\\
{\cal F}_{2D}^{\omega}\left[f_{\phi}\left(t\right)\right]({3N\Lambda^{2}}/{64d^{4}}) & d\gg\Lambda,
\end{cases} \label{corr-exp-phi2-asymp}
\end{align}
producing a crossover between $d^{-1}$ ($d^0$) and $d^{-4}$ depth dependences for ${\ell}=1$ (${\ell}=2$) as can be seen in Fig.~\ref{fig2}(e) [Fig.~\ref{fig2}(f)]. This result demonstrates that the dependence $d^{-4}$ cannot be attributed solely to the electric noise arising from fluctuating dipole charges, and that the $d$ dependence for short characteristic correlation lengths ($\Lambda$) depends strongly on the spatial form of the correlation, i.e., $e^{-|r|/\Lambda}$ or $e^{-|r|^2/\Lambda^2}$. Figures~\ref{fig2}(e) and (f) show $d^{-1}$, $d^{-2}$ and $d^{-3}$ dependences for intermediate values of $d$. Finally, for {both} ${\ell}=1$ and ${\ell}=2$ we see that the bigger (smaller) the $\Lambda$ the stronger (weaker) the suppression of the noise for $d\ll \Lambda$ ($d\gg \Lambda$).

\begin{figure}[t!]
\begin{center}
\includegraphics[clip=true,width=1\columnwidth]{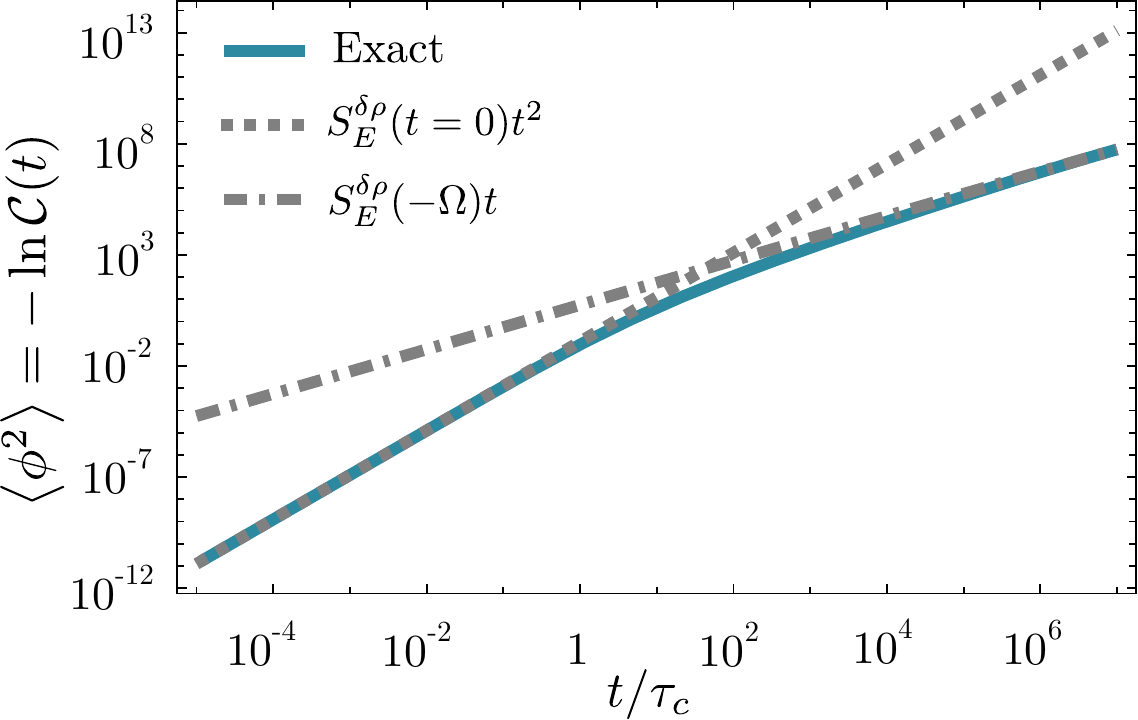}
\caption{Plot of $\left \langle \phi^2 \right \rangle= -\ln {\cal C}(t)$ [Eq.~(\ref{deco})] as a function $t/\tau_c$. The solid line represents the exact numeric evaluation of Eq.~(\ref{deco}), and the dashed and dot-dashed lines represent the asymptotic behaviors of Eq.~(\ref{deco-asymp}).} 
\label{fig3}
\end{center}
\end{figure}

{\it Sensing the diffusion with spin defects.---} Our theory suggests sensing the diffusive regime using spin centers, and extracting the corresponding diffusion constant ${\cal D}$. This could occur via multiple spin centers located at different depths. The crossover from $d^{-4}\rightarrow d^{-3} \rightarrow d^{-2} \rightarrow d^{-1} \rightarrow -\ln d$ should be seen through {both} $T_1^{-1}$ {and} $T_2^{-1}$, with experimental results showing a trend similar to Figs.~\ref{fig1} and \ref{fig2}(a) as $T_{1,2}^{-1}\propto S_E(\omega)$. From the critical depth of the crossover,  $d_c$,  the diffusion coefficient would be found through ${\cal D}=d_c^2 {\omega}$. The diffusive regime could also be identified with only one spin center located at $d$, by measuring $1/T_1$ or $1/T_2$ as a function of $\omega$, which should follow a trend similar to Fig.~\ref{fig2}(c) with a crossover located at $\omega_c$, and a diffusion coefficient of ${\cal D}=d^2{\omega_{c}}$. 

In analogy with recent work using spin centers to sense many-body interactions~\cite{davis2021probing,dwyer2021probing}, here we {define} {the average probe coherence,} ${\cal C}(t)$, {which characterizes the time evolution of both} the dephasing or relaxation processes due to  electric noise. For Gaussian noise ${\cal C}(t)=e^{-(\delta^2/2)\int_{0}^t dt' \int_{0}^{t}dt'' \left \langle E_{\mu}(t'')E_{\mu}(t') \right \rangle e^{i\Omega(t''-t')}}$~\cite{candido-pin,magneticnoise6} where $\delta$ is proportional to the {spin-defect} electric dipole moment. $\Omega$ {characterizes either spin relaxation, with its value given by the corresponding frequency difference of the levels, or spin decoherence for $\Omega=0$~\cite{candido-pin,candido-spin1surface,magneticnoise6}}. In terms of the electric noise spectral density
\begin{equation}
    {\cal C}(t) = {\rm exp}\left\{ -\frac{\delta^{2}}{2}\int_{-\infty}^{\infty}\frac{d\omega}{2\pi}S_{E}\left(\omega\right)\frac{\sin^{2}\left[\left(\omega+\Omega\right)t/2\right]}{\left[\left(\omega+\Omega\right)/2\right]^{2}}\right\}, 
    \label{deco}
\end{equation}
with a limit ${\cal C}(t)\approx \text{exp}\left[-\frac{\delta^2}{2}S_{E}(t=0)t^2\right]$ with $S_E(t=0)=\int_{-\infty}^{\infty}d\omega/(2\pi)S_{{E}}(\omega)$ for $t\ll \tau_c$, and ${\cal C}(t)\approx \text{exp}\left[-\frac{\delta^2}{2}S_{{E}}^{}(\omega=-\Omega)t\right]$ for $t\gg \tau_c$ {with ${\tau_c=d^2/{\cal D}}$}.
For diffusion the asymptotic behaviors are
\begin{equation}
    {\cal C}(t)=\begin{cases}
{\rm exp}\left[{-\frac{\delta^{2}}{2}\frac{B_0}{4d^2}t^{2}}\right] & \quad t\ll \tau_c\\
\vspace{-3mm}\\
{\rm exp}\left[-\frac{\delta^{2}}{2}S_{E}^{\delta \rho}(-\Omega)t\right] & \quad t\gg \tau_c
\end{cases},\label{deco-asymp}
\end{equation}
which indicates a crossover between two different exponential decays, namely, $e^{-At^2}$ and $e^{-Bt}$. 
As $\lim_{\omega\rightarrow 0}S_{E}^{\delta \rho}(\omega)\rightarrow \infty$, this asymptotic behavior only holds for $\Omega\neq0$. For $\Omega=0$ we still find asymptotic behavior, but it is obtained by first performing the integral of Eq.~(\ref{deco}) and then taking the limit $\Omega\rightarrow0$.
Thus  decoherence and {relaxation} dominated by carrier diffusion can be demonstrated and observed experimentally {through the time evolution of ${\cal C}(t)$, shown in Fig.~\ref{fig3}}. This makes it possible to determine the diffusion correlation time as $\tau_c=\tau_c({\cal D})$. The depth dependence $\propto d^{-2}$ is only obtained if  $t\ll \tau_c$, which causes the temporal decay of ${\cal C}(t)$ to be independent of ${\cal D}$; for short times  diffusion  does not influence the decoherence or relaxation of the spin center. For $t\gg \tau_c$ the depth dependence is dictated by  {Eq.~(\ref{SEx-rho})}.

%{\it Sensing shot noise with spin defects.---} Our theory suggests sensing shot noise  with spin centers. Since the noise of the current is also produced by the fluctuation of the charge density under  an applied electric field, i.e., $\left \langle \delta j_\nu(t') \delta j_\nu(t)\right \rangle \propto e^2\left\langle \delta n(t') \delta n(t) \right\rangle$, a clear relationship between the shot noise and $T_{1,2}$ can be established, although this is beyond the scope of this work. 

{\it Concluding remarks.---} We have developed a theory which predicts that the depth dependence of spin center decoherence and decay processes are strongly influenced by both the two-point correlation of the fluctuating surface particles' density and the surface electrostatic potential. Our framework predicts that diffusive phenomena yield a non-trivial temporal behavior for the average probe coherence of spin centers, with a crossover between different exponential decay forms determined by the effective correlation time at the spin center. As a consequence, both the crossover point and the exponents contain a direct signature of diffusion, and permit  extraction of the diffusion coefficient. %Finally, our theory is general and permits to probe and understand surface correlating phenomena beyond the analyzed diffusion case, including sensing of shot noise and hydrodynamics.}

\begin{acknowledgments}
We thank D. D. Awschalom, P. C. Maurer, M. Fukami, and J. C. Karsch for useful discussions. The work was supported as part of the Center for Molecular Quantum Transduction, an Energy Frontier Research Center funded by the U.S. Department of Energy, Office of Science, Basic Energy Sciences, under Award Number DE-SC0021314.
\end{acknowledgments}

\bibliography{main.bib}

\end{document}